# Data-Driven Insights into Rare Earth Mineralization: Machine Learning Applications Using Functional Material Synthesis Data

Juejing Liu,[1,2,#] Xiaoxu Li,[1,#] Yifu Feng,[1] Zheming Wang,[1] Kevin M. Rosso,[1] Xiaofeng Guo,[2] and Xin Zhang[1,*]

1. Physical & Computational Science Directorate, Pacific Northwest National Laboratory, Richland, Washington 99354, United States
2. Department of Chemistry, Washington State University, Pullman, Washington 99164, United States

[#]These authors contribute equal.

***Corresponding Authors:** xin.zhang@pnnl.gov (X.Z.)

**Abstract**

Improving our understanding of rare earth element (REE) mineralization mechanisms in natural systems could be useful for developing new REE separation strategies on an industrial scale. However, although qualitatively much is known about geochemical processes giving rise to REE deposits, there is a lack of quantitative data on specific conditions leading to specific mineralization outcomes. In this study, we adapted the laboratory REE hydrothermal synthesis data for functional material fabrication as a surrogate to study REE mineralization with a data-driven method. From previous such studies we built a REE crystallization dataset of more than 1200 hydrothermal reaction data points. Three ML models (KNN, RF, and XGB) were trained to predict product elements and phase from precursor, additive, reaction conditions, and various augmented features. New experiments were performed to test model validity. The XGB model exhibited the highest accuracy in predicting both product elements and phase. Feature importance analysis showed that the XGB model heavily relied on the thermodynamic properties of cation/anion to make predictions. Correlation analysis revealed a positive correlation among reaction time, precursor concentration, pH, and reaction temperature, which aligns with the classical crystallization mechanism for REE minerals. The comparison between our experimental results of synthesizing REE minerals and model prediction shows that the model relies on similar examples in the training dataset to make predictions of the phase and product, which is expected in ML approaches. We further trained XGB-based models to successfully predict the crystallization temperature and pH from the precursor and product. This study shows that it is viable to use data abstracted from one discipline to make predictions in that of another, by demonstrating that material science data can be used to understand natural REE mineralization. It also highlights the importance of investigating the crystallization of less-studied REE minerals, e.g., REE carbonates and phases involving heavy REEs. This approach has the potential to accelerate both the development of more efficient REE extraction and separation methods as well as the discovery of new REE deposits.

## Introduction

Rare earth elements (REEs) are critical materials for a wide range of industries due to their unique chemical and physical properties. These elements play a crucial role in modern technology and sustainable energy solutions. For example, REEs are used in catalysts for petroleum refining and pollution control, in permanent magnets for electric vehicles and wind turbines, in phosphors for lighting and displays, and in nuclear reactors for control rods and shielding.[1-6] Despite their significant industrial value, the availability of REEs is constrained by the geographic distribution of economically viable deposits.[7-9] Additionally, the extraction of REEs is concentrated in a limited number of mines, primarily located in China, Australia, and the United States.[9] To address potential supply risks, researchers are exploring diverse REE sources beyond conventional mining of primary deposits. These include processing of other mine tailings as well as alternative sources such as industrial waste, electronic waste (e-waste), and coal fly ash. [9-11] Developing these non-traditional sources could help stabilize the REE market by diversifying the supply chain.[12] The key steps in REE extraction are the enrichment and separation of REEs from each other, a process that remains a significant technical challenge.[13] The current state-of-the-art separation technique relies on organic ligands,[13, 14] which selectively extracts REEs from complex mixtures in solution. While this technique is highly effective, its high cost and environmental risk have limited its application.[15] Therefore, the development of alternative separation techniques is necessary to accommodate a broader range of REE minerals and unconventional feedstocks.

In primary deposits, REE's often exhibit separation through natural mineralization processes.[15, 16] Therefore, studying enrichment and separation during mineralization under different geochemical conditions may help develop alternative re-processing strategies. Understanding the natural processes that lead to REE deposit formation can also provide insights into their distribution in various environmental settings.[17-19] Hence such knowledge can help both in the search for new REE deposits and in improving extraction methods. The key to advancing this understanding lies in comprehensively understanding the mineralization processes of REEs, particularly in hydrothermal environments, where many REE deposits are formed.

However, while the geochemical processes leading to most known REE deposits are reasonably well understood, datasets containing detailed information on REE mineralization reactions are lacking.[20-22] Geological surveys provide valuable data on various REE-bearing minerals, but they often lack insights into the specific mineralization reactions occurring under different geochemical conditions. Currently, datasets focused explicitly on REE mineralization mechanisms are scarce, limiting researchers' ability to analyze and develop more generalizable predictive models for REE formation and behavior. To overcome this challenge, one promising approach involves leveraging data from laboratory-based synthesis of REE-based functional materials. Many REE-based materials, such as photoluminescent compounds, catalyst matrices, and solid-state lasers, are synthesized using hydrothermal methods that closely mimic natural REE mineralization conditions[5, 23-26]. By analyzing these synthesis processes and outcomes, in principle researchers can gain valuable insights into REE mineralization mechanisms valuable both for developing separation strategies as well as understanding REE formation in geological settings.[27] One major advantage of this approach is the abundance of available data from REE-based functional materials research. The rapid growth of this field has led to extensive datasets detailing the synthesis conditions, precursor compositions, and resulting phases of various REE materials.[28] While this approach shows promise, the datasets are highly multi-modal and multi-condition, such that the most useful analysis strategy is not immediately clear. This is particularly true given that one of the objectives is to seek generalizable insights that span from laboratory synthesis to conditions and reaction mechanisms relevant to those in natural hydrothermal environments. A robust analysis would thus require systematic comparisons between experimentally synthesized REE materials and naturally occurring REE minerals.[29]

To investigate the viability of using REE functional material synthesis data as a surrogate for REE geochemical reactions, we trained machine learning (ML) models on the synthesis data and applied them to geochemical-mimicking reactions. ML models readily learn and generalize the relationships among different features and labels.[30-32] The viability of this approach was then investigated by evaluating the model's performance in terms of accuracy in predicting reaction outcomes. Moreover, by analyzing the impact of input features on the predicted labels, it is possible to gain insights into the reaction mechanisms of REE geochemical reactions.[33-35] We demonstrate that this approach can help predict REE mineral formation in geochemical settings and provide insights into the underlying mineralization mechanisms.

We collected 1239 REE hydrothermal reaction data points from functional material synthesis literature and organized them into a dataset.[28] ML models were trained using three algorithms: K-nearest neighbor (KNN),[36] random forest (RF),[37] and extreme gradient boost (XGB).[38] These models aimed to predict the elements presented in the product minerals and phases in the product from precursors, reaction conditions, and augmented features, such as ionic radii and electronegativity. The XGB-based model exhibited the highest accuracy in predicting both product elements and phase, as measured by F1 score. We conducted feature importance analysis and correlation analysis on the model and dataset to investigate relationships between features and labels, such as relationships between precursor concentration and product phase, as well as correlations between temperature and pH. We discovered that the model relied heavily on the standard formation enthalpy and entropy at 25 °C and 1 bar of the species in solution to predict phase and apparent elements. We then experimentally evaluated the model's accuracy by synthesizing REE minerals under randomly generated conditions with an interpolation approach. The parameters we generated were within the range of the data. We then compared the experimental results to the model's predictions. The result shows that the model heavily relies on the examples in the training dataset to make predictions. When similar examples are presented during training, the model correctly predicts the experimental result. In contrast, when similar examples are limited (especially for REE carbonate), the prediction accuracy deteriorated. We further trained XGB-models to predict the crystallization pH and temperature for REE minerals. This study demonstrates the potential of using REE functional material synthesis data to mimic REE mineralization in geochemical settings. However, more data is needed from less-investigated REE minerals, such as REE carbonates and heavy REE minerals, to improve the model's generalizability.

## Methods and Experiments

**Data collection and building dataset.** We systematically collected REE hydrothermal reaction data from published papers reporting the synthesis of REE functional materials.[28] The resulting dataset comprises 1239 REE material synthesis reactions. For each reaction, the dataset includes the chemical formula and concentration of all precursors and additives, reaction conditions (pH, temperature, time, solvent type and concentration), and product properties (chemical formula and phase). We augmented the dataset with additional features derived from various sources, including elemental properties (Mendeleev package),[39] anion/cation thermodynamic properties (CHNOSZ package),[40] formation energy of reactants (ML models),[41-43] solution properties,[44] and additive properties (Bordwell pKa Table).[45] The dataset was randomly split into training (80%) and validation (20%) sets.

**Model training and validation.** We trained three different machine learning models—K-nearest neighbor (KNN), random forest (RF), and extreme gradient boost (XGB)—to predict the product elements and phase from the precursor, additive, and reaction conditions using the scikit-learn library.[46] The models were trained as multi-label classifiers to predict the presence or absence of each element and phase in the product. Cross-entropy was used as the loss function to optimize the models. Five-fold cross-validation was used to evaluate the models' performance during training. The models output a binary vector, where each element represents the presence (1) or absence (0) of a specific element or phase in the product. All input features were normalized using the StandardScaler function in scikit-learn to ensure that features with different scales did not disproportionately influence the models. We performed hyperparameter tuning using GridSearchCV to find the optimal set of hyperparameters for each model. The range of hyperparameters explored and the best set of hyperparameters for each model are listed in **Table S1**. The trained models were then evaluated on the held-out validation dataset to assess their generalization performance. We used confusion matrices to visualize the models' performance in terms of true positives, true negatives, false positives, and false negatives. Labels that were consistently negative across both the predicted and ground truth results were excluded from the confusion matrix calculations to avoid inflating the accuracy metrics.

**Feature importance and correlation analysis.** We performed a feature importance analysis using the XGB model. Due to the high dimensionality of our dataset, we used an alternative approach to permutation feature importance analysis, which would have been computationally expensive. Our approach involved training a series of XGB models, each with one type of feature removed. We then evaluated the importance of each feature by comparing the prediction accuracy of the reduced model (with the feature removed) to the accuracy of the original model (with all features included). The decrease in accuracy was normalized to a scale of 0 to 1, where 0 represents no change in accuracy and 1 represents the most significant decrease in accuracy (see **Table S2**). We also performed Pearson correlation analysis to investigate the relationships between reaction conditions (precursor concentrations, time, pH, temperature, and volume of solvent).[47]

**Synthesis of REE minerals.** All chemicals, including $Nd(NO_3)_3\cdot6H_2O$ (>99.9%), $Er(NO_3)_3\cdot5H_2O$ (>99.9%), $ErCl_3\cdot6H_2O$ (>99.9%), $La(NO_3)_3\cdot6H_2O$ (>99.9%), $LaCl_3\cdot7H_2O$ (ACS reagent), $Gd(NO_3)_3\cdot6H_2O$ (>99.9%), $GdCl_3\cdot6H_2O$ (>99%), $Na_2CO_3$ (>99.5%), and $Na_2HPO_4$ (>99.0%) were purchased from Sigma-Aldrich and used as received. REE carbonate samples were synthesized via a co-precipitation method. In a typical procedure, a 0.01 M rare earth nitrate ($REE(NO_3)_3$) solution was mixed with a 0.05 M sodium carbonate ($Na_2CO_3$) solution at room temperature. The pH was adjusted to the desired range (9.00–12.35) using 5 M NaOH, followed by heating at different temperatures (25°C–175°C) in 22 mL Parr vessels with a Teflon liner inside an electrical oven. The total reaction volume was maintained at 15 mL, and the reaction proceeded for 46 hours. For REE phosphate synthesis, a 0.032 M $REECl_3$ solution was mixed with an equimolar 0.032 M sodium hydrogen phosphate ($Na_2HPO_4$) solution. The pH was adjusted to the desired values (1.52–10.70) using 5 M HCl or 5 M NaOH. The solution was then heated at different temperatures in a Parr vessel with a Teflon liner inside an oven for 48 hours. The resulting precipitate washed with deionized water three times, separated by centrifuge and dried for subsequent analysis. The detailed experimental conditions are listed in **Table S3** and **S4**.

**Prediction of crystallization temperature and pH.** To predict the crystallization temperature and pH, we trained an XGB regression model (see **Table S5** for hyperparameters). The model used precursors, additives, augmented features, and products as input features to predict the crystallization temperature and pH. We normalized the input features using the StandardScaler function and performed hyperparameter tuning using GridSearchCV to optimize the model's performance. Mean squared error (MSE) was used as the loss function. We evaluated the model's performance on the validation dataset using the coefficient of determination ($r^2$).

## Results

We trained three types of machine learning (ML) models—K-nearest neighbor (KNN), random forest (RF), and extreme gradient boost (XGB)—to predict the REE elements present in the product and the phase of the crystalline sample, using the type and concentration of reactants and additives, solvent conditions (e.g., pH, type, and volume), and reaction parameters (mainly temperature and time) as input features (**Figure 1** and **Table S1**).[28] The dataset was further augmented with properties from elements in reactant, cation and anions in solution, additive, and solvent (**Figure 1a**, also see method section for details). 80% of the data was used for training with five-fold cross-validation, and the remaining 20% was used for testing. The augmented dataset was fed into three algorithms: K-nearest neighbor (KNN, **Figure 1b**), random forest (RF, **Figure 1c**), and eXtreme Gradient Boosting (XGB, **Figure 1d**) to study the trends and patterns involved in the reactions.[36-38] KNN is a clustering-based algorithm that learns common features from different groups in the dataset. When presented with a new sample, the model compares its features to the common features of different clusters and classifies it into the cluster with the highest similarity. RF uses multiple decision trees to make predictions collectively. During training, the criteria in each tree are tuned, and when a new sample is presented, it is classified by every tree simultaneously. The final prediction is the averaged result from all the trees. XGB, a gradient boosting algorithm, also uses decision trees but in a sequential manner. It starts with an initial tree and then iteratively adds new trees to correct the errors of the previous ones. This process continues until a certain limit is reached or the prediction accuracy no longer improves. This sequential approach makes XGB particularly well-suited for handling complex datasets.

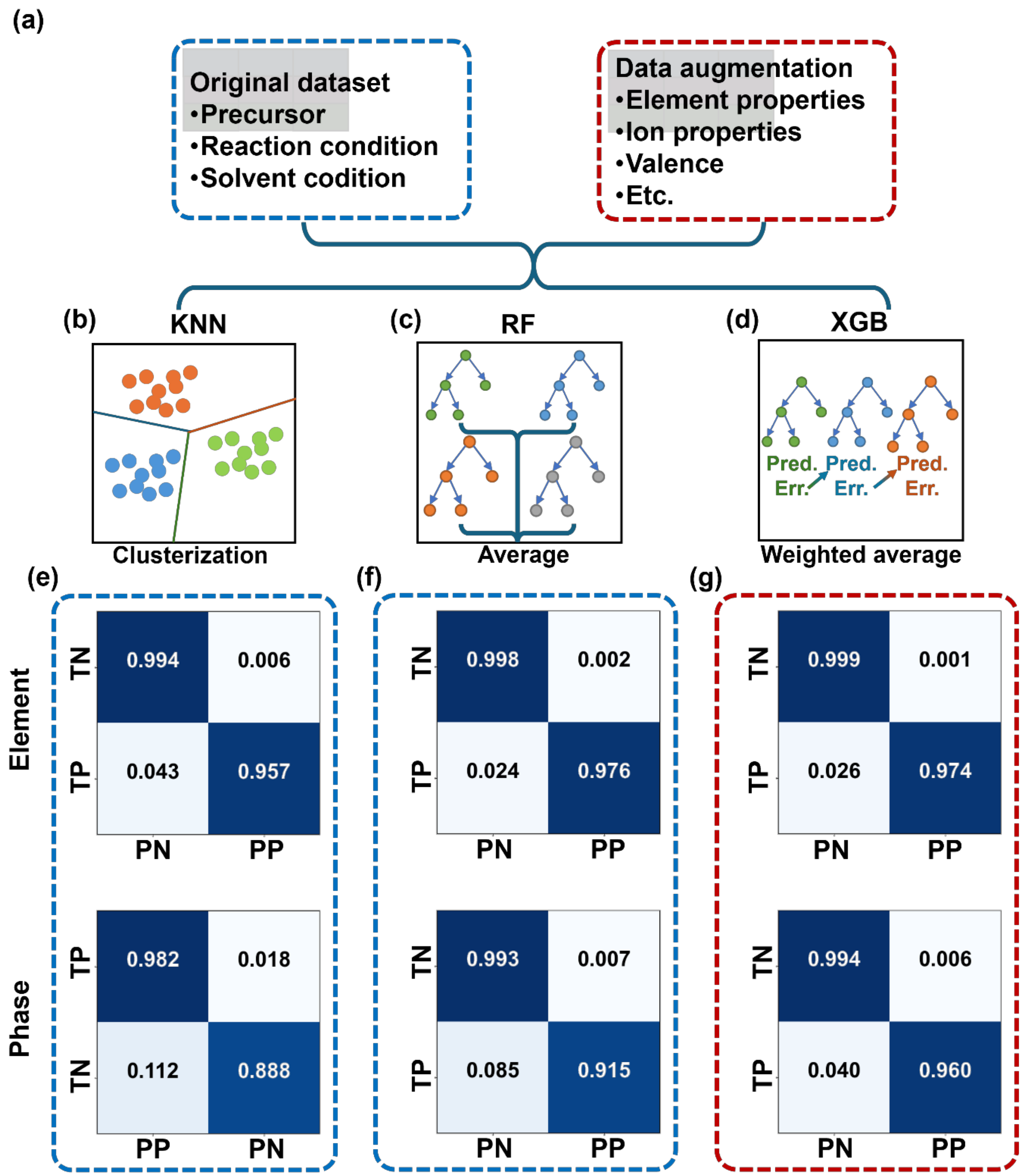

**Figure 1**. Three machine learning algorithms applied to predict the result of REE crystallization reactions. (a) Input features in dataset and from augmentation. (b) to (d) Illustration of three algorithms used in learning trends of REE crystallization, including K nearest neighbor (KNN, b), random forest (RF, c), and eXtreme Gradient Boosting (XGB, d). (e) to (g) Confusion matrix of prediction results for the models from the three algorithms, including KNN (e), RF (f), and XGB (g). TP, TN, PP, and PN correspond to true positive, true negative, predicted positive, and predicted negative.

The XGB-based model achieved the highest accuracy among the three models, particularly in predicting phases from the reactions. We used normalized confusion matrices to visualize the performance of the models for both element and phase prediction. All models exhibited high accuracy (above 90%) in predicting negative results, such as elements and phases not presented in the product. For predicting the presence of specific elements in the product, the KNN model exhibited 95.7% accuracy. The RF and XGB models showed slightly increased accuracy in element prediction, with correct positive prediction rates of 97.6% and 97.4%, respectively. This superior performance with a wide variety of features demonstrates that the RF and XGB models are better suited for handling complex datasets compared to the KNN model.

The XGB model also excelled in predicting the formation of phases from the experimental parameters. The correct positive prediction rate for phases was 88.8% for the KNN model, 91.5% for the RF model, and 96.0% for the XGB model. The superior performance of the XGB model can be attributed to its unique sequential architecture, where each new tree corrects the errors of the previous trees, allowing it to effectively learn complex relationships within the data. Given its superior prediction accuracy, we selected the XGB model for further analysis and used it to predict the elements and phases present in unseen experiments.

We performed a feature importance analysis to investigate the relationship between different features and prediction accuracy. Permutation feature importance analysis was initially considered but deemed computationally impractical due to the large number of features in our dataset.[48] Therefore, we developed an alternative method involving training a series of XGB models, each with one type of input feature removed. We then compared the training accuracy of each reduced model to the original model with all features, using cross-entropy validation. The decrease in accuracy was normalized by setting the maximum decrease to 1.00 and the accuracy of the original model to 0.00.

As shown in **Figure 2a** and **Table S2**, the feature importance analysis considered three main groups of features: augmented features, reaction parameters, and concentration of reactants and additives. The augmented features encompass various physicochemical properties of the elements, ions, and compounds involved in the reactions. These features were derived from external databases and literature sources to provide additional information beyond the basic reaction parameters. Examples of augmented features include ion thermodynamic properties (e.g., enthalpy and entropy of formation), ion mass, reactant formation energy, valence of elements in reactants and ions, properties of reactant elements (e.g., electronegativity, atomic radius), additive properties (e.g., pKa values), ion properties (e.g., ionic radius), and solvent properties (e.g., polarity).

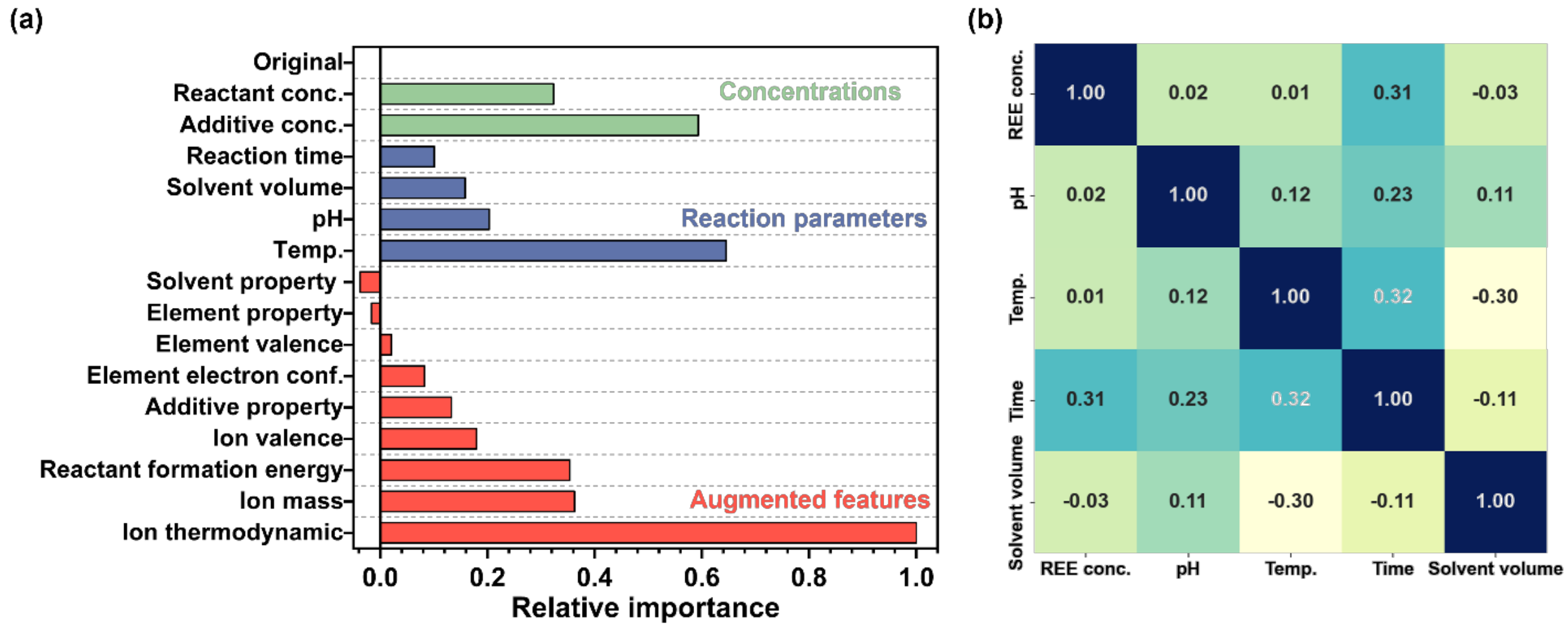

**Figure 2.** Feature importance analysis for XGB model (a) and correlation analysis on major experimental parameters (b).

The second group, reaction parameters, includes the experimental conditions under which the reactions were carried out, such as temperature, time, pH, and solvent volume. These features are directly controllable and have a significant impact on the reaction outcome. The final group, concentration of reactants and additives, represents the amounts of reactants and additives used in the reactions, expressed in moles per liter of solution (mol/L).

The thermodynamic properties of ions, obtained from the CHNOSZ package.[40] emerged as the most important augmented feature, with a relative importance of 1.00. This highlights the crucial role of thermodynamic properties in governing crystallization reactions. Similarly, the formation energy of reactants (0.35), obtained from ML-based calculations,[41-43] was also identified as an important feature, further emphasizing the influence of thermodynamic factors on the reaction outcome. Interestingly, the ion mass (0.36) was also identified as a highly important feature. This could be because the ion mass serves as a unique identifier for different ions, as the masses of different ions are rarely identical.

The analysis also revealed that ion valences (0.18) are more important than element valences (0.02) in predicting the reaction outcome. This is likely because the reactions primarily involve the recombination of ions, where the valence of the element remains unchanged. The additive properties (0.13) also play a role in determining the reaction outcome. These properties include the pKa values of ionic additives and the critical micelle concentration (CMC) for surfactant-type additives. The importance of pKa values in predicting the product element and phase is consistent with previous findings that the pKa of additives influences the crystallization of REE minerals.[49] However, the exact mechanism by which pKa affects REE mineral precipitation and crystallization remains unclear. Other features, such as electron configuration (0.08) of elements, atomic properties (-0.01, including atomic mass, electronegativity, etc.), and solvent properties (-0.04, polarity), were found to be less important in predicting the reaction outcome. The limited importance of these features is likely because the elements and solvents themselves are not directly involved in the reaction.

Among the reaction parameters, temperature (0.64) emerged as the most important feature. This is consistent with the well-established role of temperature in determining the phase of REE minerals formed in hydrothermal reactions.[50, 51] The pH (0.20) was also identified as an important feature, likely because different reactions involving REE cations and various anions (e.g., carbonate, phosphate, hydroxide) exhibit distinct pH preferences. Solvent volume (0.16) and reaction time (0.10) were also found to influence the prediction of reaction products. Reaction time can influence the phase of the product, particularly for REE-phosphate reactions.[52] Since our dataset only includes REE hydrothermal reactions conducted in autoclaves, the solvent volume likely affects the pressure inside the reactor, which in turn can influence the crystallization of REE minerals.[31] Regarding concentrations, the additive concentration (0.59) was found to be more important than the reactant concentration (0.32). This suggests that the outcome of REE mineralization, particularly the phase of the product, is more sensitive to the concentration of additives than to the concentration of reactants.[49]

We further analyzed the correlations among reaction parameters (temperature, time, pH, and solvent volume) and reactant and additive concentrations using the Pearson correlation coefficient (see **Figure 2b**).[47] We observed positive correlations between reaction time and pH (0.23), temperature (0.32), and REE concentration (0.31). These positive correlations suggest that well-crystallized REE minerals are typically obtained under conditions of high reaction temperature, long reaction time, and high pH, while poorly crystallized products are associated with low temperature, short reaction time, and low pH. Interestingly, we also observed a significant negative correlation between temperature and solvent volume (-0.30). This negative correlation could be attributed to the limitations of autoclaves, as the combination of high temperature and large solvent volume can lead to high pressure buildup and potential leakage. This suggests that researchers tend to avoid such conditions to ensure the safety and integrity of their experiments.

**Table 1.** Summary of REE phosphate experimental results from randomly generated conditions vs. prediction from XGB model. See **Figure S1-S3** and **Table S3** for details.

| No. | Precursor | Exp. result | | Prediction | |
|---|---|---|---|---|---|
| | | Element | Phase | Element | Phase |
| 1 | $LaCl_3$-$Na_2HPO_4$ | La, P, O | Hexagonal | La, P, O | Hexagonal |
| 2 | $LaCl_3$-$Na_2HPO_4$ | La, P, O | Hexagonal | La, P, O | Hexagonal |
| 3 | $LaCl_3$-$Na_2HPO_4$ | La, P, O | Hexagonal | La, P, O | Hexagonal |
| 4 | $GdCl_3$- $Na_2HPO_4$ | Gd, P, O | Hexagonal | Gd, P, O | Hexagonal |
| 5 | $GdCl_3$- $Na_2HPO_4$ | Gd, P, O | Hexagonal | Gd, P, O | Hexagonal |
| 6 | $GdCl_3$- $Na_2HPO_4$ | Gd, P, O | Hexagonal | Gd, P, O | Hexagonal |
| 7 | $GdCl_3$- $Na_2HPO_4$ | Gd, P, O | Hexagonal | Gd, P, O | Hexagonal |
| 8 | $GdCl_3$- $Na_2HPO_4$ | Gd, P, O | Hexagonal | Gd, P, O | Hexagonal |
| 9 | $GdCl_3$- $Na_2HPO_4$ | Gd, P, O | Hexagonal | Gd, P, O | Hexagonal |
| 10 | $ErCl_3$- $Na_2HPO_4$ | Er, P, O | Tetragonal | Er, P, O | - |
| 11 | $ErCl_3$- $Na_2HPO_4$ | Er, P, O | Tetragonal | Er, P, O | - |
| 12 | $ErCl_3$- $Na_2HPO_4$ | Er, P, O | Tetragonal | Er, P, O | - |
| 13 | $ErCl_3$- $Na_2HPO_4$ | Er, P, O | Tetragonal | Er, P, O | - |
| 14 | $ErCl_3$- $Na_2HPO_4$ | Er, P, O | Tetragonal | Er, P, O | - |
| 15 | $ErCl_3$- $Na_2HPO_4$ | Er, P, O | Tetragonal | Er, P, O | - |

To further validate the XGB model, we conducted a series of hydrothermal synthesis experiments to synthesize REE phosphates and carbonates under conditions not included in the training or validation datasets (see **Tables 1, 2, S3,** and **S4** for experimental details). We compared the experimental results to the model predictions to assess the model's ability to generalize to unseen conditions. As expected, the model's prediction accuracy was influenced by the presence of similar reactions in the training dataset. In the phosphate synthesis experiments, the XGB model accurately predicted the presence of La, Gd, and Er in the products, as well as the hexagonal phase of the La and Gd phosphates. However, the model incorrectly predicted the phase of the Er phosphate. This misclassification could be attributed to the underrepresentation of heavy REE phosphate synthesis reactions in the training dataset.[28] Despite the limited number of heavy REE phosphate examples, the model was able to correctly predict the presence of Er in the product. However, the model's performance was less consistent for the REE carbonate synthesis experiments.

**Table 2.** Summary of REE carbonate experimental results from randomly generated conditions vs. prediction from XGB model. See **Figure S4-S7** and **Table S4** for details.

| No. | Precursor | Exp. result | | Prediction | |
|---|---|---|---|---|---|
| | | Element | Phase | Element | Phase |
| 1 | $La(NO_3)_3$-$Na_2CO_3$ | La, H, C, O | Orthorhombic | La, C, O | Monoclinic |
| 2 | $La(NO_3)_3$-$Na_2CO_3$ | La, H, C, O | Orthorhombic | La, C, O | Monoclinic |
| 3 | $La(NO_3)_3$-$Na_2CO_3$ | La, H, C, O | Orthorhombic | La, O | - |
| 4 | $La(NO_3)_3$-$Na_2CO_3$ | La, H, C, O | Amorphous | La, C, O | - |
| 5 | $La(NO_3)_3$-$Na_2CO_3$ | La, H, C, O | Amorphous | La, C, O | - |
| 6 | $La(NO_3)_3$-$Na_2CO_3$ | La, H, C, O | Orthorhombic | La, O | - |
| 7 | $La(NO_3)_3$-$Na_2CO_3$ | La, H, C, O | Amorphous | La, C, O | Monoclinic |
| 8 | $La(NO_3)_3$-$Na_2CO_3$ | La, H, C, O | Amorphous | La, C, O | Monoclinic |
| 9 | $La(NO_3)_3$-$Na_2CO_3$ | La, H, C, O | Hexagonal | La, O | - |
| 10 | $Er(NO_3)_3$-$Na_2CO_3$ | Er, H, C, O | Amorphous | Er, C, O | - |
| 11 | $Er(NO_3)_3$-$Na_2CO_3$ | Er, H, C, O | Amorphous | Er, O | - |
| 12 | $Er(NO_3)_3$-$Na_2CO_3$ | Er, H, C, O | Amorphous | Er, O | - |
| 13 | $Er(NO_3)_3$-$Na_2CO_3$ | Er, H, C, O | Amorphous | Er, C, O | - |
| 14 | $Er(NO_3)_3$-$Na_2CO_3$ | Er, H, C, O | Amorphous | Er, O | - |
| 15 | $Er(NO_3)_3$-$Na_2CO_3$ | Er, H, C, O | Amorphous | Er, O | - |
| 16 | $Er(NO_3)_3$-$Na_2CO_3$ | Er, H, C, O | Amorphous | Er, C, O | - |
| 17 | $Er(NO_3)_3$-$Na_2CO_3$ | Er, H, C, O | Amorphous | Er, O | - |
| 18 | $Er(NO_3)_3$-$Na_2CO_3$ | Er, H, C, O | Amorphous | Er, O | - |
| 19 | $Nd(NO_3)_3$-$Na_2CO_3$ | Nd, H, C, O | Amorphous | C, O | - |
| 20 | $Nd(NO_3)_3$-$Na_2CO_3$ | Nd, H, C, O | Amorphous | C, O | - |
| 21 | $Nd(NO_3)_3$-$Na_2CO_3$ | Nd, H, C, O | Hexagonal | C, O | - |
| 22 | $Nd(NO_3)_3$-$Na_2CO_3$ | Nd, H, C, O | Amorphous | C, O | - |
| 23 | $Nd(NO_3)_3$-$Na_2CO_3$ | Nd, H, C, O | Amorphous | C, O | - |
| 24 | $Nd(NO_3)_3$-$Na_2CO_3$ | Nd, H, C, O | Hexagonal | C, O | - |
| 25 | $Nd(NO_3)_3$-$Na_2CO_3$ | Nd, H, C, O | Orthorhombic | C, O | - |
| 26 | $Nd(NO_3)_3$-$Na_2CO_3$ | Nd, H, C, O | Amorphous | C, O | - |
| 27 | $Nd(NO_3)_3$-$Na_2CO_3$ | Nd, H, C, O | Hexagonal | C, O | - |
| 28 | $Gd(NO_3)_3$-$Na_2CO_3$ | Gd, H, C, O | Amorphous | Gd, C, O | orthorhombic |
| 29 | $Gd(NO_3)_3$-$Na_2CO_3$ | Gd, H, C, O | Amorphous | Gd, C, O | orthorhombic |
| 30 | $Gd(NO_3)_3$-$Na_2CO_3$ | Gd, H, C, O | Amorphous | Gd, C, O | - |
| 31 | $Gd(NO_3)_3$-$Na_2CO_3$ | Gd, H, C, O | Amorphous | Gd, C, O | orthorhombic |
| 32 | $Gd(NO_3)_3$-$Na_2CO_3$ | Gd, H, C, O | Amorphous | Gd, C, O | orthorhombic |
| 33 | $Gd(NO_3)_3$-$Na_2CO_3$ | Gd, H, C, O | Hexagonal | Gd, C, O | - |
| 34 | $Gd(NO_3)_3$-$Na_2CO_3$ | Gd, H, C, O | Amorphous | Gd, C, O | - |
| 35 | $Gd(NO_3)_3$-$Na_2CO_3$ | Gd, H, C, O | Amorphous | Gd, C, O | - |
| 36 | $Gd(NO_3)_3$-$Na_2CO_3$ | Gd, H, C, O | Amorphous | Gd, C, O | - |

The XGB model's performance on the REE carbonate synthesis experiments was limited by the scarcity of similar reactions in the training dataset and the inherent differences between functional material synthesis and geochemical mineralization. These reactions typically produced REE carbonates with varying hydration levels or hydroxycarbonates, resulting in complex product phases (orthorhombic, hexagonal, amorphous) containing REE, C, H, and O. The scarcity of REE carbonate examples in the training dataset had a greater impact on phase prediction than on element prediction, as the model failed to accurately predict any of the product phases.

In the La and Er carbonate synthesis experiments, the model's element prediction accuracy was inconsistent. For La carbonate, the model correctly identified La, C, and O in 6 out of 9 experiments, while in the remaining 3 experiments, it only identified La and O. The Er carbonate predictions followed a similar pattern but with lower accuracy, correctly identifying Er, C, and O in only 3 out of 9 experiments. In the remaining 6 Er carbonate experiments, the model only predicted the presence of Er and O. The model performed worst for the Nd carbonate synthesis, failing to identify Nd in any of the products. In contrast, the model accurately predicted the presence of Gd, C, and O in all Gd carbonate synthesis experiments.

To investigate the model's failure to predict Nd-carbonate reactions, we examined the training dataset, particularly the reactions involving Nd. We found that while the dataset contains Nd carbonate synthesis reactions, these reactions used urea as the carbon source instead of carbonate salts, which are more relevant to geochemical conditions. This difference in reaction conditions likely explains the model's inability to accurately predict Nd-carbonate mineralization under geochemical conditions where carbonate ions are prevalent.

Temperature and pH are among the key reaction conditions influencing the crystallization of REE minerals. Because different REE minerals often have distinct crystallization conditions, controlling these conditions (such as temperature and pH) can enable the separation of different REEs from a solution. Accurately predicting the crystallization conditions for specific REE minerals is thus crucial for developing effective REE separation strategies. To explore this, we trained an XGB regression model to predict the crystallization temperature and pH, using reactant, product, and augmented features as input (**Figure 3a** and **3b**, respectively). Despite the limited size of the dataset, the XGB regression model demonstrated promising predictive ability. The coefficients of determination ($r^2$) for both temperature and pH prediction were 0.87, indicating a good fit between the predicted and actual values. This result demonstrates the potential for predicting REE separation parameters using ML models trained on reaction datasets. To further enhance the accuracy of these predictions, expanding the dataset with more reaction examples is necessary.

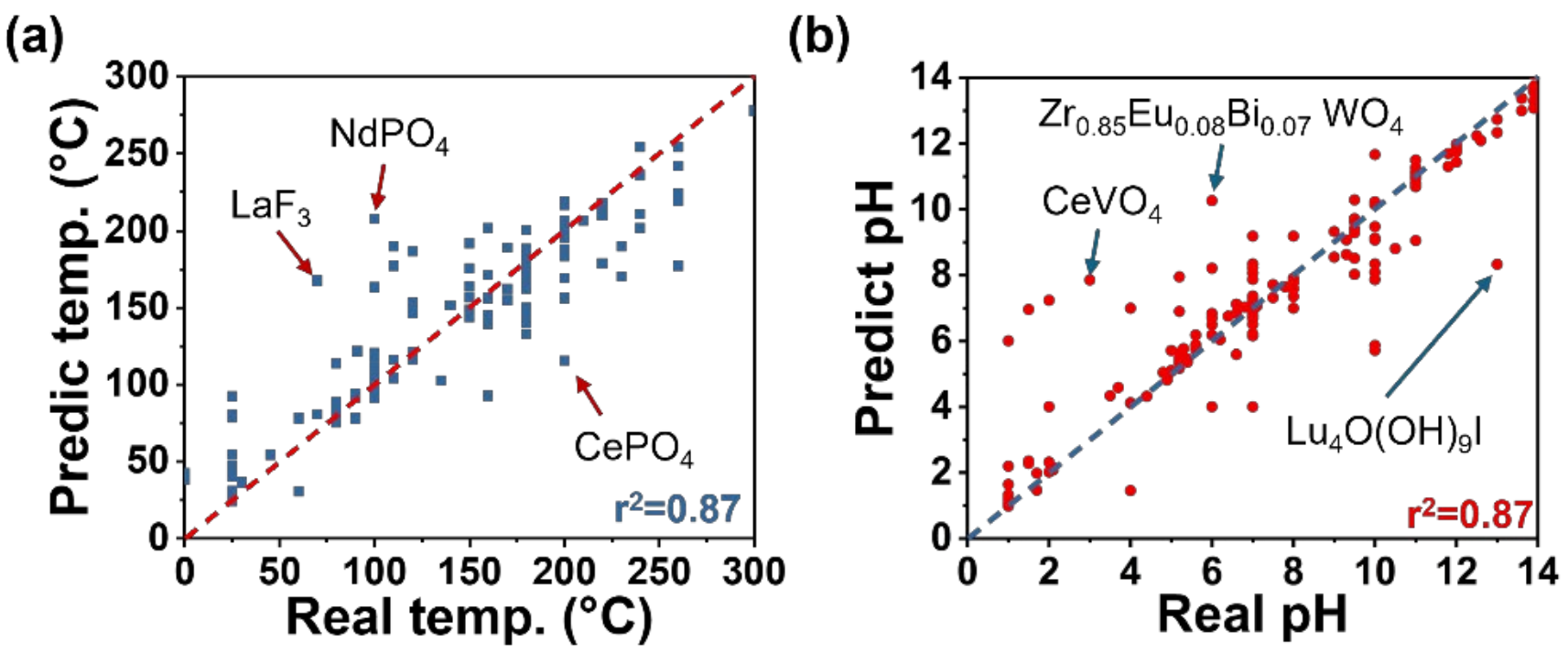

**Figure 3.** Prediction of crystallization temperature (a) and pH (b) using reactant, additive, and product as input.

Although both temperature and pH predictions showed similar overall accuracy ($r^2 = 0.87$), examining individual predictions with significant errors revealed differences in the nature of these errors. In the temperature prediction (**Figure 3a**), the largest errors were associated with relatively common minerals, such as fluorite and phosphate. Since these anions frequently combine with REE cations in nature,[53, 54] incorporating more reaction examples involving these minerals could potentially improve the accuracy of temperature prediction. In contrast, the pH predictions with the largest errors involved products containing anions like vanadate, tungstate, and the complex anion $O(OH)_9I^{12-}$, which are less common in natural REE minerals (**Figure 3b**). The scarcity of these anions in natural REE minerals limits the availability of relevant reaction examples, which may explain the higher prediction errors for pH. To improve the accuracy of pH prediction, further experimental studies focusing on the crystallization of REE minerals with these less common anions are needed.

## Discussion

This study aimed to investigate the feasibility of using data from one domain (e.g., synthesis of REE minerals for practical applications) to study another similar domain (e.g., REE mineralization). This approach was motivated by the limited availability of data on REE mineral geochemical reactions with clear statement of precursor and reaction conditions. Our results demonstrate that this is a viable approach, although some challenges and limitations remain. We trained various ML models to predict reaction products from precursors, additives, and reaction conditions and found that the XGB model performed best. The feature importance analysis revealed the relative importance of different features in predicting the reaction outcome, while the correlation analysis highlighted relationships between reaction parameters.

These findings provide valuable insights into the factors that influence REE mineral crystallization and can guide further investigations into the underlying mechanisms.

We observed that the model's ability to generalize to unseen experimental data depends on the availability of similar examples in the training dataset. For reactions with sufficient representation in the training dataset, such as those between light and medium REEs with phosphate, the model accurately predicted both the elements and phases of the product (e.g., monazites type REE phosphate). As we limited the range of randomly generated parameters aligning with the training dataset, out model relies on interpolation to predict the result. However, for reactions with limited representation in the training dataset, such as those involving heavy REEs (e.g., xenotime type REE phosphate) or carbonate anions, the model struggled to predict the correct product properties, particularly the phase. This limitation aligns with the properties of interpolation models, where a comprehensive dataset is key to train models to make correct predictions.

The observed limitations in the model's performance for certain reactions appear to stem from an unbalanced dataset, prompting further analysis of the reasons behind this imbalance. Our dataset was primarily constructed from reactions reported in studies focused on synthesizing REE mineral particles for various applications, such as photoluminescent materials, sensors, and other functional material.[23, 55] This focus on practical applications has led to a bias in the dataset, with certain REE minerals, particularly phosphates, being more frequently reported than others, such as carbonates. This lack of data for less-studied reactions makes it challenging for ML models to identify the underlying patterns and trends. Our findings underscore the importance of expanding research efforts to include the mineralization of REEs with less-studied anions, such as carbonates and those involving heavy REEs. Data from such studies will be crucial for developing more comprehensive and accurate ML models for predicting REE mineralization, ultimately enhancing our understanding of REE geochemistry.

## Conclusion

In this study, we developed ML models to predict the products of REE crystallization reactions using a dataset compiled from functional material synthesis literature. The XGB model achieved the highest accuracy, particularly when predicting the phase of the products. Our analysis revealed the importance of thermodynamic properties, reaction conditions, and additive concentrations in influencing REE crystallization. The comparison between model prediction and our experiment results shows that the model is capable of predicting the REE mineralization product when similar examples are available in the training dataset, e.g., phosphate. In contrast, the model is not able to predict the REE mineralization when the similar examples are limited, such as carbonate. We also demonstrated the potential for predicting crystallization temperature and pH using an XGB regression model. This study establishes the feasibility of using functional material synthesis data to study REE mineralization, offering a valuable approach to address the scarcity of traditional REE mineralization data. Our findings underscore the need for further research on a wider range of REE minerals, which will contribute to the development of more robust predictive models and a deeper understanding of REE geochemistry. This knowledge will ultimately benefit both fundamental research and the development of sustainable REE resource management strategies.

**Data Availability**

We have made the data publicly available. The iron oxide synthesis dataset is available at https://data.pnl.gov/XX.

**Code Availability**

The code that can be used to replicate the results presented in this work is available by request.

**Acknowledgments**

All authors acknowledge support from the Advanced Research Projects Agency-Energy's (ARPA-E) Mining Innovations for Negative Emissions Resource Recovery (MINER) program with award number 0002707-1515, "Re-Mining Red Mud Waste for $CO_2$ Capture and Storage and Critical Element Recovery (RMCCS-CER)". J. L. and X. Z. also acknowledge support from a Laboratory Directed Research and Development Project at Pacific Northwest National Laboratory (PNNL). K. M. R. also acknowledges support from the U.S. Department of Energy, Office of Science, Office of Basic Energy Sciences, Chemical Sciences, Geosciences, and Biosciences Division through its Geosciences program at Pacific Northwest National Laboratory (PNNL) (FWP 56674). X.G. also acknowledges the support of this work by the National Science Foundation (NSF), Division of Earth Sciences, under award No. 2149848. A portion of the work was carried out in the Environmental and Molecular Sciences Laboratory (EMSL), a national scientific user facility at PNNL sponsored by the DOE Office of Biological and Environmental Research, under user proposals 51382 (10.46936/lser.proj.2020.51382/60000186) and 61223 (10.46936/lsr.proj.2024.61223/60012698). PNNL is a multi-program national laboratory operated by Battelle Memorial Institute under contract no. DE-AC05-76RL01830 for the DOE.

**AUTHOR CONTRIBUTIONS**

X. Z. conceived the project. X. Z., X. G., and K.M.R. supervised the project. J. L. collected the literature data, developed the database, performed the data analysis, and conducted the ML. X. L. performed the experiments works and analyzed the experimental data with the help from Y. F. J. L., X. L., and X. Z. wrote the manuscript with inputs from all co-authors. J. L. and X. L. are considered the co-first author. Z.W., X.

G. and K. M. R. helped with project design and manuscript refinement. All authors have given approval to the final version of the manuscript.

## Ethics declarations

The authors declare no competing financial interests.

# Supplementary Information of

# Data-Driven Insights into Rare Earth Mineralization: Machine Learning Applications Using Functional Material Synthesis Data

Juejing Liu,[1,2,#] Xiaoxu Li,[1,#] Yifu Feng,[1] Zheming Wang,[1] Kevin M. Rosso,[1] Xiaofeng Guo,[2] and Xin Zhang[1,*]

1. Physical & Computational Science Directorate, Pacific Northwest National Laboratory, Richland, Washington 99354, United States
2. Department of Chemistry, Washington State University, Pullman, Washington 99164, United States

[#]These authors contribute equal.

***Corresponding Authors:** xin.zhang@pnnl.gov (X.Z.)

Table S1. Training parameters ranges for GirdsearchCV function used in training KNN, RF, and XGB models. The best parameters are also listed.

| Algorithm | Parameters | Best parameter |
|---|---|---|
| KNN | n_neighbors: 1, 2, 3, 5 | 1 |
| | base_estimator__leaf_size: 10, 20, 30, 40 | 10 |
| | estimator__p: 1, 2, 3 | 1 |
| RF | n_estimators: 400, 1000, 1200, 1500 | 1200 |
| | max_depth: 100, 200, 400, 600 | 100 |
| | min_samples_split: 2, 5, 7 | 2 |
| XGB | n_estimators: 1200, 1500, 2000 | 1200 |
| | max_depth: 1, 2 | 2 |
| | Eta: 0.2, 0.3, 0.4, 0.5 | 0.2 |

Table S2. Accuracy of models after removing specific feature and the corresponding importance values.

| Removed Feature | Accuracy | Importance value |
|---|---|---|
| Ion thermodynamic | 0.88751 | 1 |
| Ion mass | 0.89585 | 0.36264 |
| Reactant formation energy | 0.89597 | 0.3535 |
| Ion valence | 0.89825 | 0.17947 |
| Additive property | 0.89886 | 0.13252 |
| Element electron conf. | 0.89952 | 0.08232 |
| Element valence | 0.90033 | 0.02063 |
| Element property | 0.90082 | -0.01651 |
| Solvent property | 0.90109 | -0.03778 |
| Temp. | 0.89215 | 0.64514 |
| pH | 0.89794 | 0.20321 |
| Solvent volume | 0.89852 | 0.1584 |
| Reaction time | 0.89928 | 0.10073 |
| Additive conc. | 0.89283 | 0.59324 |
| Reactant conc. | 0.89637 | 0.32333 |
| Original | 0.9006 | 0 |

Table S3. Experimental conditions for synthesizing rare earth element phosphates.

| No. | precursor 1 (P1) | concentration of P1 (mol/L) | precursor 2 (P2) | concentration of P2 (mol/L) | pH | Temp (°C) | Time (hour) | Volume (ml) |
|---|---|---|---|---|---|---|---|---|
| 1 | $LaCl_3$ | 0.032 | $Na_2HPO_4$ | 0.032 | 5.83 | 180 | 48 | 15 |
| 2 | $LaCl_3$ | 0.032 | $Na_2HPO_4$ | 0.032 | 5.83 | 80 | 48 | 15 |
| 3 | $LaCl_3$ | 0.032 | $Na_2HPO_4$ | 0.032 | 10.7 | 180 | 48 | 15 |
| 4 | $GdCl_3$ | 0.032 | $Na_2HPO_4$ | 0.032 | 1.52 | 180 | 48 | 15 |
| 5 | $GdCl_3$ | 0.032 | $Na_2HPO_4$ | 0.032 | 1.52 | 80 | 48 | 15 |
| 6 | $GdCl_3$ | 0.032 | $Na_2HPO_4$ | 0.032 | 5.32 | 180 | 48 | 15 |
| 7 | $GdCl_3$ | 0.032 | $Na_2HPO_4$ | 0.032 | 5.32 | 80 | 48 | 15 |
| 8 | $GdCl_3$ | 0.032 | $Na_2HPO_4$ | 0.032 | 10.84 | 180 | 48 | 15 |
| 9 | $GdCl_3$ | 0.032 | $Na_2HPO_4$ | 0.032 | 10.84 | 80 | 48 | 15 |
| 10 | $ErCl_3$ | 0.032 | $Na_2HPO_4$ | 0.032 | 1.84 | 180 | 48 | 15 |
| 11 | $ErCl_3$ | 0.032 | $Na_2HPO_4$ | 0.032 | 1.84 | 80 | 48 | 15 |
| 12 | $ErCl_3$ | 0.032 | $Na_2HPO_4$ | 0.032 | 5.69 | 180 | 48 | 15 |
| 13 | $ErCl_3$ | 0.032 | $Na_2HPO_4$ | 0.032 | 5.69 | 80 | 48 | 15 |
| 14 | $ErCl_3$ | 0.032 | $Na_2HPO_4$ | 0.032 | 10.9 | 180 | 48 | 15 |
| 15 | $ErCl_3$ | 0.032 | $Na_2HPO_4$ | 0.032 | 10.9 | 80 | 48 | 15 |

Table S4. Experimental conditions for synthesizing rare earth element carbonates.

| No. | precursor 1 (P1) | concentration of P1 (mol/L) | precursor 2 (P2) | concentration of P2 (mol/L) | pH | Temp (°C) | Time (hour) | Volume (ml) |
|---|---|---|---|---|---|---|---|---|
| 1 | $La(NO_3)_3$ | 0.01 | $Na_2CO_3$ | 0.05 | 12.35 | 25 | 46 | 15 |
| 2 | $La(NO_3)_3$ | 0.01 | $Na_2CO_3$ | 0.05 | 12.35 | 80 | 46 | 15 |
| 3 | $La(NO_3)_3$ | 0.01 | $Na_2CO_3$ | 0.05 | 12.35 | 175 | 46 | 15 |
| 4 | $La(NO_3)_3$ | 0.01 | $Na_2CO_3$ | 0.05 | 9 | 25 | 46 | 15 |
| 5 | $La(NO_3)_3$ | 0.01 | $Na_2CO_3$ | 0.05 | 9 | 80 | 46 | 15 |
| 6 | $La(NO_3)_3$ | 0.01 | $Na_2CO_3$ | 0.05 | 9 | 175 | 46 | 15 |
| 7 | $La(NO_3)_3$ | 0.01 | $Na_2CO_3$ | 0.05 | 7 | 25 | 46 | 15 |
| 8 | $La(NO_3)_3$ | 0.01 | $Na_2CO_3$ | 0.05 | 7 | 80 | 46 | 15 |
| 9 | $La(NO_3)_3$ | 0.01 | $Na_2CO_3$ | 0.05 | 7 | 175 | 46 | 15 |
| 10 | $Er(NO_3)_3$ | 0.02 | $Na_2CO_3$ | 0.05 | 12 | 25 | 46 | 15 |
| 11 | $Er(NO_3)_3$ | 0.02 | $Na_2CO_3$ | 0.05 | 12 | 80 | 46 | 15 |
| 12 | $Er(NO_3)_3$ | 0.02 | $Na_2CO_3$ | 0.05 | 12 | 175 | 46 | 15 |
| 13 | $Er(NO_3)_3$ | 0.02 | $Na_2CO_3$ | 0.05 | 9.62 | 25 | 46 | 15 |
| 14 | $Er(NO_3)_3$ | 0.02 | $Na_2CO_3$ | 0.05 | 9.62 | 80 | 46 | 15 |
| 15 | $Er(NO_3)_3$ | 0.02 | $Na_2CO_3$ | 0.05 | 9.62 | 175 | 46 | 15 |
| 16 | $Er(NO_3)_3$ | 0.02 | $Na_2CO_3$ | 0.05 | 7 | 25 | 46 | 15 |
| 17 | $Er(NO_3)_3$ | 0.02 | $Na_2CO_3$ | 0.05 | 7 | 80 | 46 | 15 |
| 18 | $Er(NO_3)_3$ | 0.02 | $Na_2CO_3$ | 0.05 | 7 | 175 | 46 | 15 |
| 19 | $Nd(NO_3)_3$ | 0.01 | $Na_2CO_3$ | 0.05 | 12 | 25 | 46 | 15 |
| 20 | $Nd(NO_3)_3$ | 0.01 | $Na_2CO_3$ | 0.05 | 12 | 80 | 46 | 15 |
| 21 | $Nd(NO_3)_3$ | 0.01 | $Na_2CO_3$ | 0.05 | 12 | 175 | 46 | 15 |
| 22 | $Nd(NO_3)_3$ | 0.01 | $Na_2CO_3$ | 0.05 | 9 | 25 | 46 | 15 |
| 23 | $Nd(NO_3)_3$ | 0.01 | $Na_2CO_3$ | 0.05 | 9 | 80 | 46 | 15 |
| 24 | $Nd(NO_3)_3$ | 0.01 | $Na_2CO_3$ | 0.05 | 9 | 175 | 46 | 15 |
| 25 | $Nd(NO_3)_3$ | 0.01 | $Na_2CO_3$ | 0.05 | 7 | 25 | 46 | 15 |
| 26 | $Nd(NO_3)_3$ | 0.01 | $Na_2CO_3$ | 0.05 | 7 | 80 | 46 | 15 |
| 27 | $Nd(NO_3)_3$ | 0.01 | $Na_2CO_3$ | 0.05 | 7 | 175 | 46 | 15 |
| 28 | $Gd(NO_3)_3$ | 0.02 | $Na_2CO_3$ | 0.05 | 12.14 | 25 | 46 | 15 |
| 29 | $Gd(NO_3)_3$ | 0.02 | $Na_2CO_3$ | 0.05 | 12.14 | 80 | 46 | 15 |
| 30 | $Gd(NO_3)_3$ | 0.02 | $Na_2CO_3$ | 0.05 | 12.14 | 175 | 46 | 15 |
| 31 | $Gd(NO_3)_3$ | 0.02 | $Na_2CO_3$ | 0.05 | 9 | 25 | 46 | 15 |
| 32 | $Gd(NO_3)_3$ | 0.02 | $Na_2CO_3$ | 0.05 | 9 | 80 | 46 | 15 |
| 33 | $Gd(NO_3)_3$ | 0.02 | $Na_2CO_3$ | 0.05 | 9 | 175 | 46 | 15 |
| 34 | $Gd(NO_3)_3$ | 0.02 | $Na_2CO_3$ | 0.05 | 7 | 25 | 46 | 15 |
| 35 | $Gd(NO_3)_3$ | 0.02 | $Na_2CO_3$ | 0.05 | 7 | 80 | 46 | 15 |
| 36 | $Gd(NO_3)_3$ | 0.02 | $Na_2CO_3$ | 0.05 | 7 | 175 | 46 | 15 |

Table S5. Training parameters for XGB model predicting crystallization temperature and pH.

| Parameters | Best parameter |
|---|---|
| n_estimators: 50, 100, 300, 500, 800 | 1200 |
| max_depth: 1, 2, 4, 6 | 2 |
| eta: 0.01, 0.05, 0.1, 0.2 | 0.2 |

`

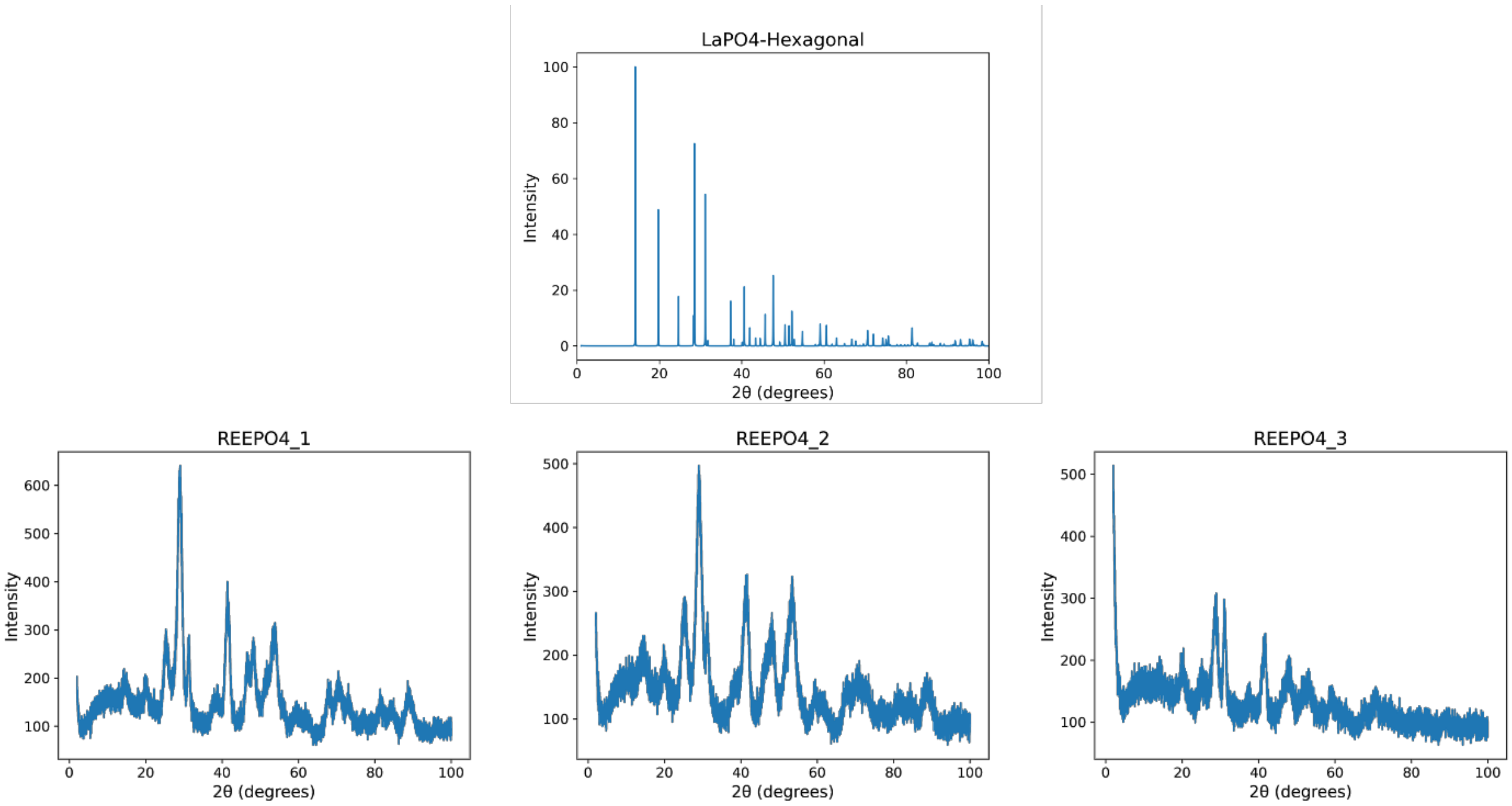

Figure S1. XRD analysis for synthesized La phosphate. The title of the plot shows the number of the sample corresponding to the synthesis conditions (see Table S3). The theoretical XRD patterns are also included.

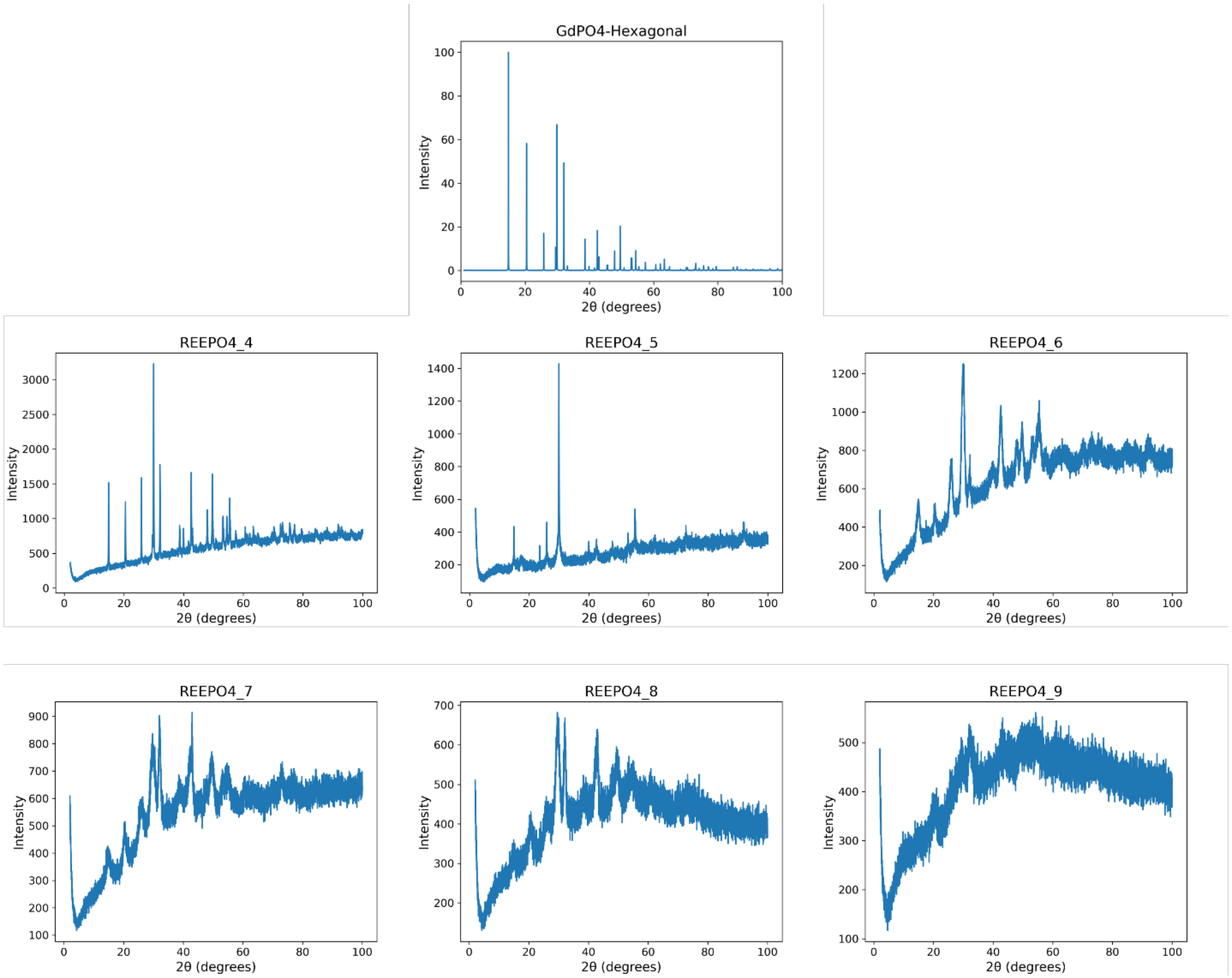

Figure S2. XRD analysis for synthesized Gd phosphate. The title of the plot shows the number of the sample corresponding to the synthesis conditions (see Table S3). The theoretical XRD patterns are also included.

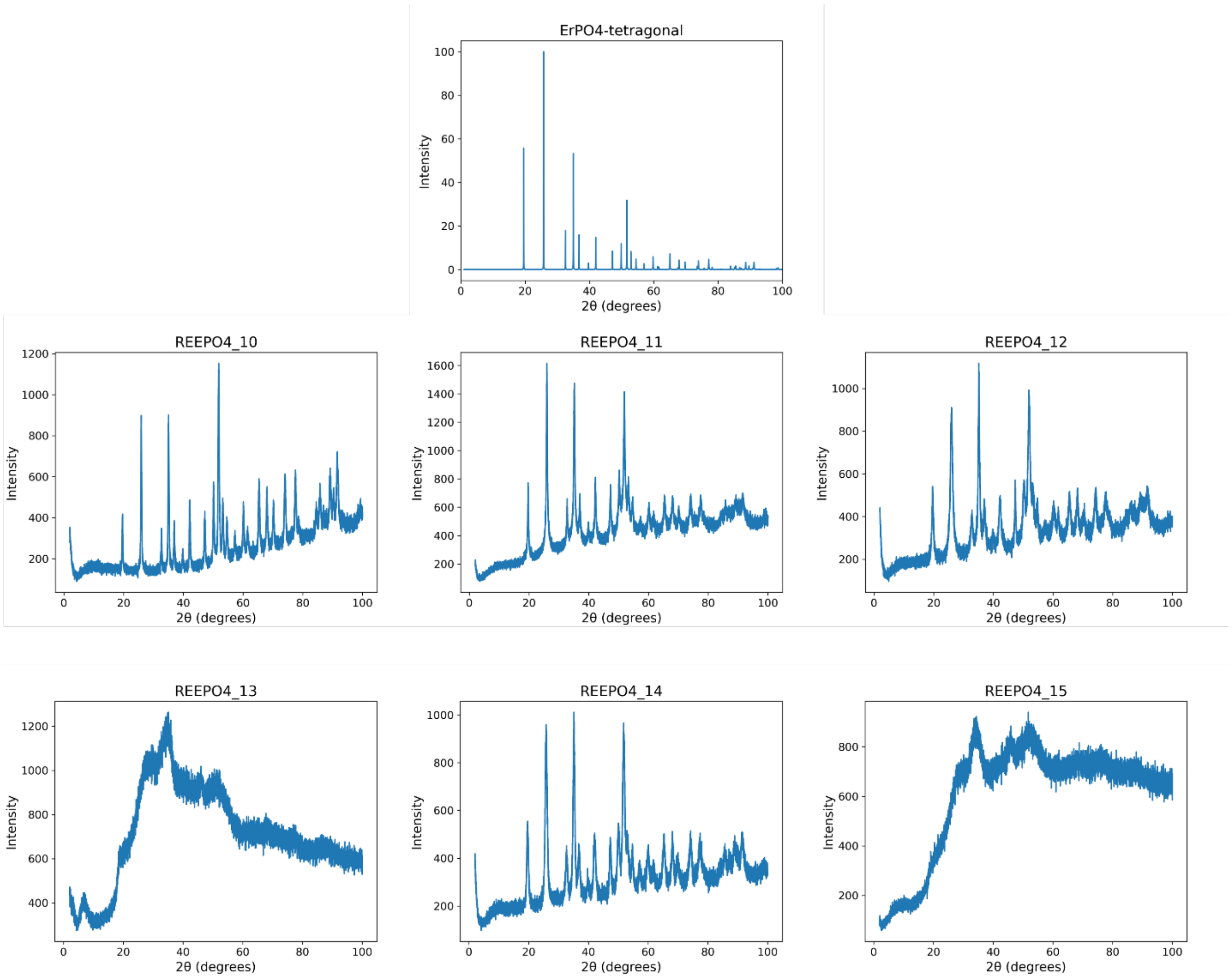

Figure S3. XRD analysis for synthesized Er phosphate. The title of the plot shows the number of the sample corresponding to the synthesis conditions (see Table S3). The theoretical XRD patterns are also included.

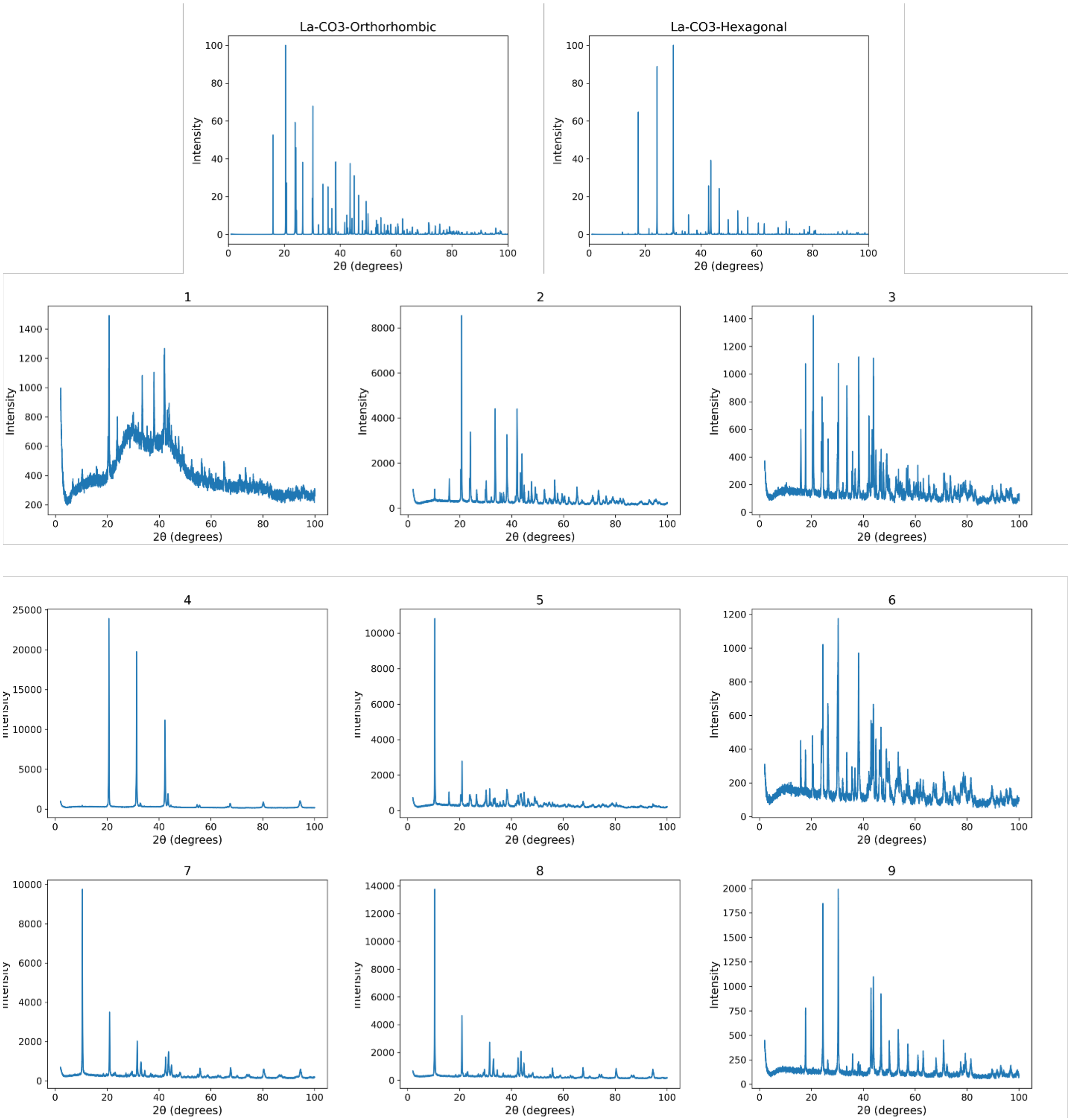

Figure S4. XRD analysis for synthesized La carbonate. The title of the plot shows the number of the sample corresponding to the synthesis conditions (see Table S4). The theoretical XRD patterns are also included.

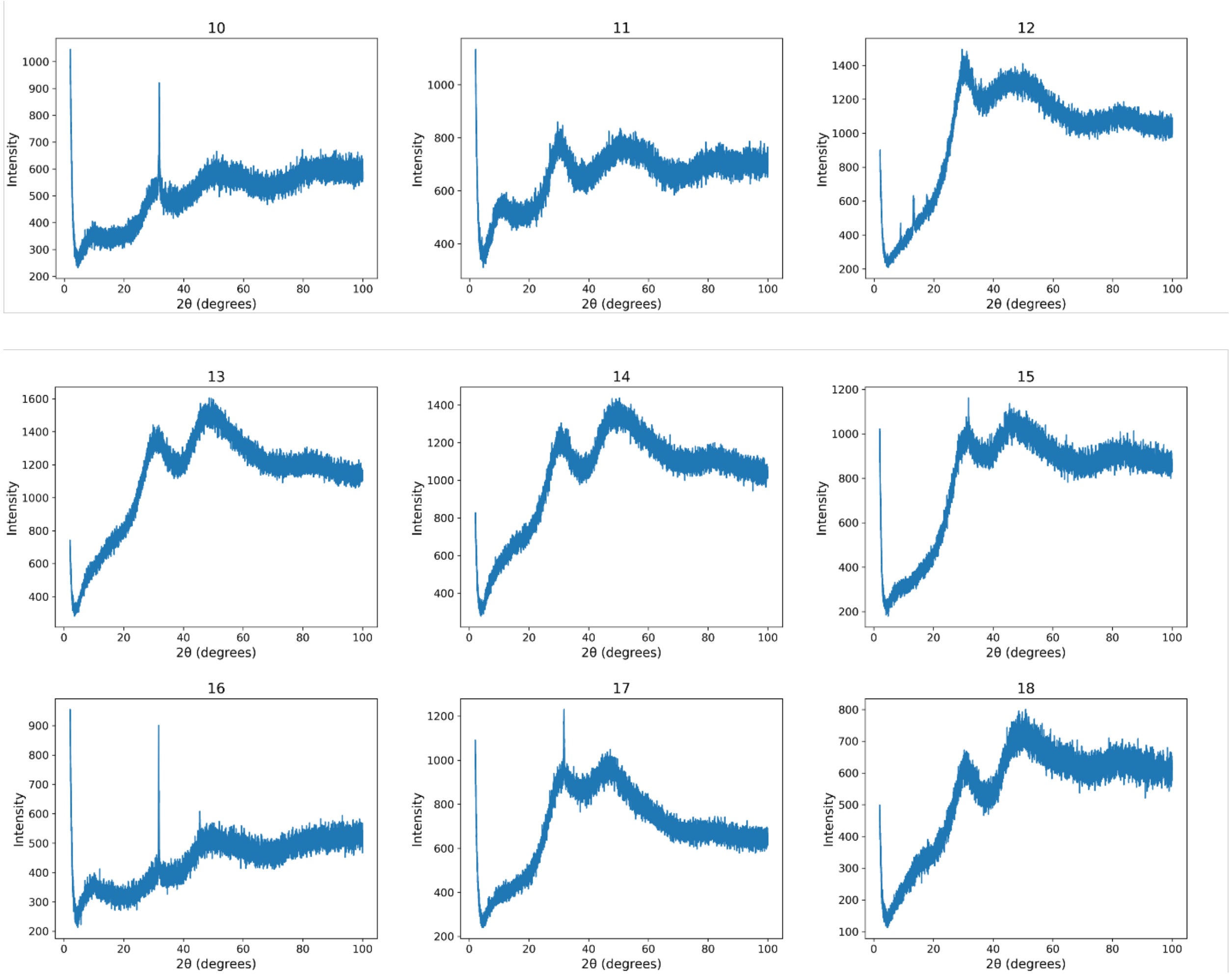

Figure S5. XRD analysis for synthesized La carbonate. The title of the plot shows the number of the sample corresponding to the synthesis conditions (see Table S4).

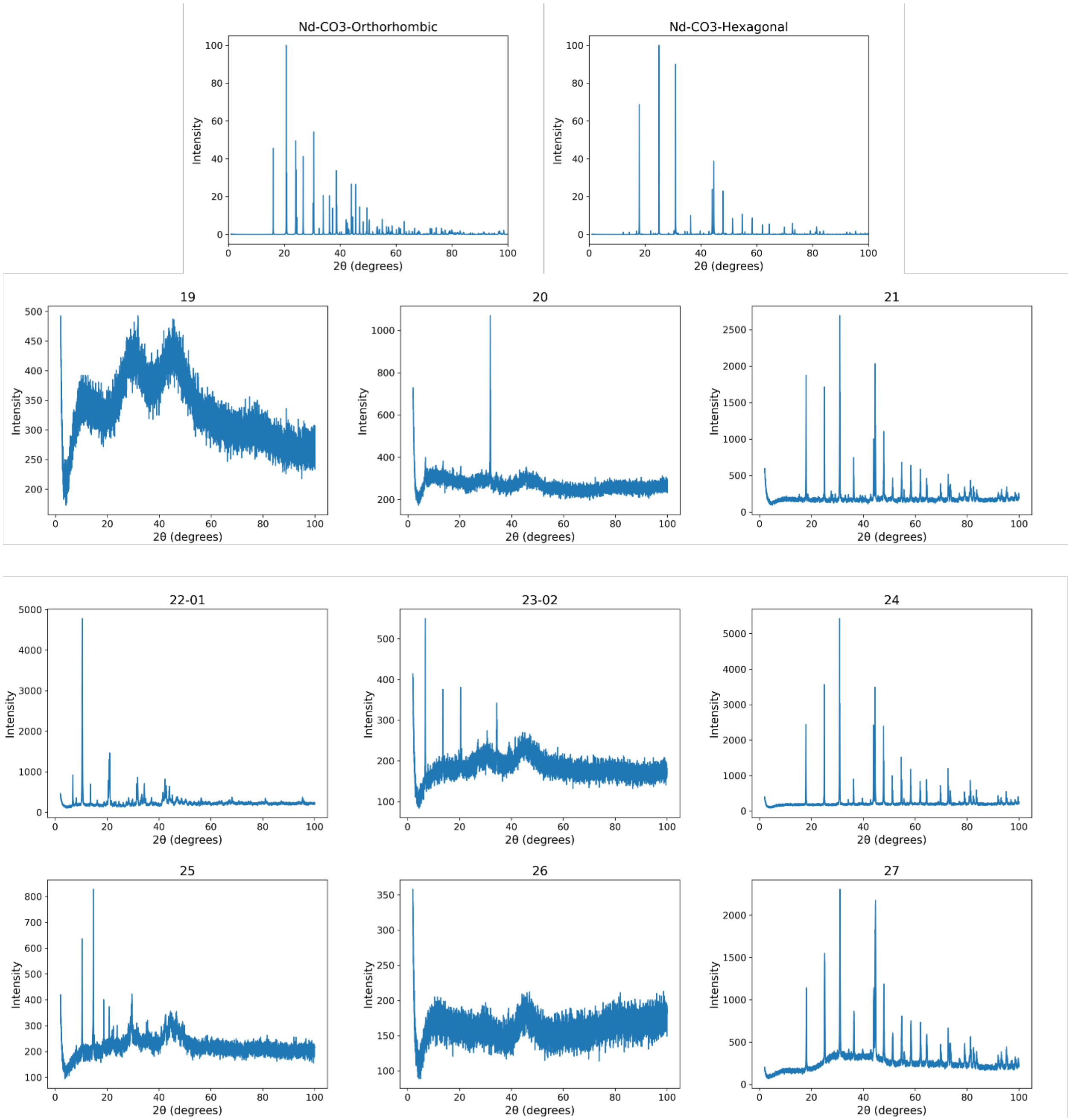

Figure S6. XRD analysis for synthesized Nd carbonate. The title of the plot shows the number of the sample corresponding to the synthesis conditions (see Table S4). The theoretical XRD patterns are also included.

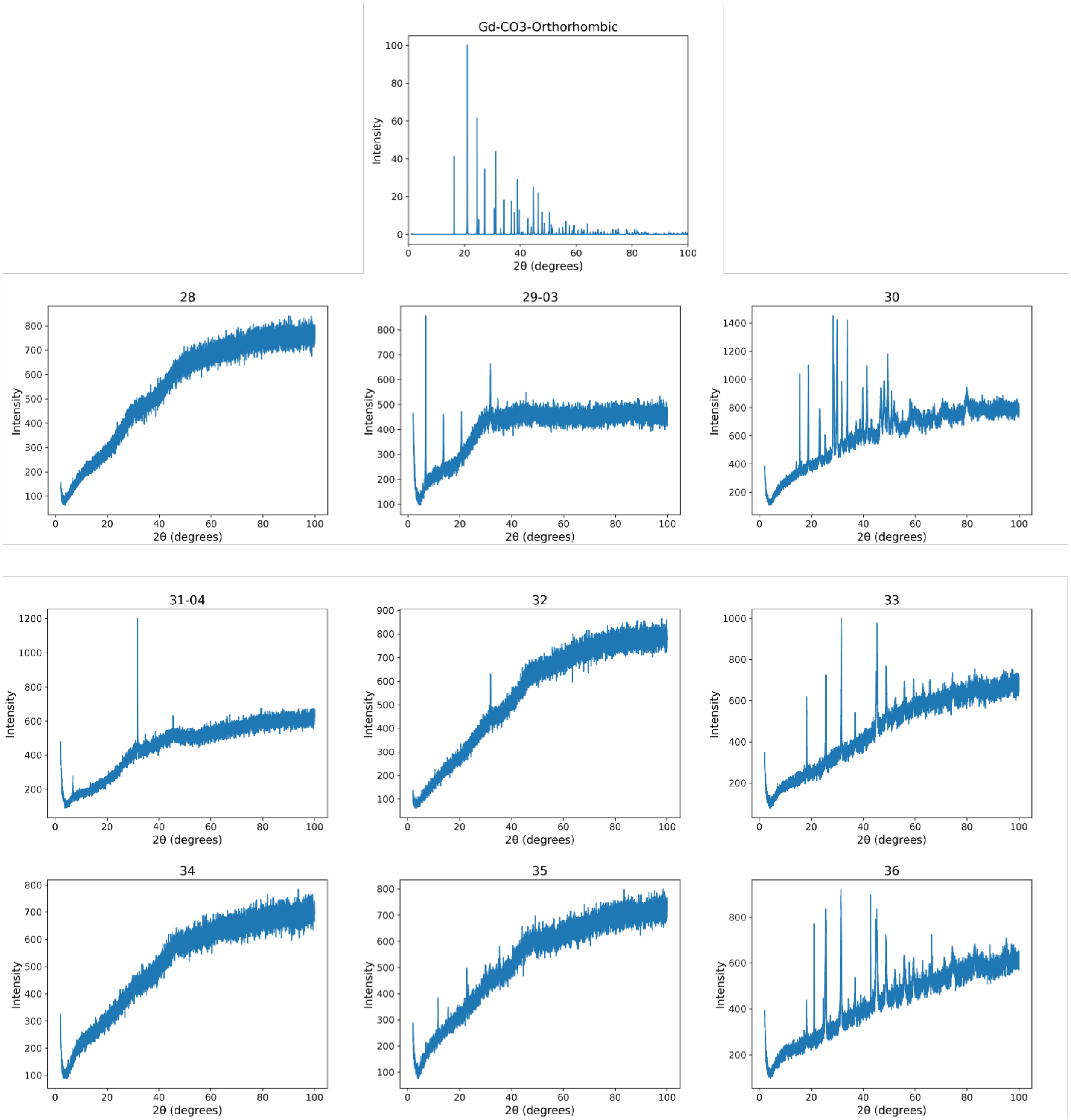

Figure S7. XRD analysis for synthesized Gd carbonate. The title of the plot shows the number of the sample corresponding to the synthesis conditions (see Table S4). The theoretical XRD patterns are also included.